\input amstex
\magnification=1200
\documentstyle{amsppt}
\NoRunningHeads
\pagewidth{142 mm}
\vcorrection{-15 mm}
\topmatter
\title
BLP dissipative structures in plane
\endtitle
\author
A.V. Yurov
\endauthor
\address {\rm 236041, Theoretical Physics Department, Kaliningrad State
University, Al.Nevsky St., 14, Kaliningrad, Russia}
\endaddress
\email
yurov\@ freemail.ru
\endemail
\abstract
We study the Darboux and Laplace transformations for the
Boiti-Leon-Pempinelli equations (BLP). These equations are the (1+2)
generalization of the sinh-Gordon equation. In addition, the BLP equations
reduced to the	Burgers (and anti-Burgers) equation in a one-dimensional limit.
Localized nonsingular solutions in both spatial
dimensions and	(anti) "blow-up" solutions are constructed. The Burgers
equation's "dressing" procedure is suggested. This procedure allows us
to construct  such solutions of the BLP equations which are reduced to the
solutions of the  dissipative Burgers equations when  $t\to \infty$.
These solutions we
call the  BLP  dissipative structures.
\endabstract
\endtopmatter
\document
{\bf 1. INTRODUCTION}
\newline
\newline

Localized structures are frequently associated with dispersion and
nonlinearity. A large class of nonlinear evolution equations with
dispersion in two spatial dimensions have been studied. Some of them are
physically relevent
equations, for example the Davey-Stewartson (DS) and
the Kadomtsev-Petviashvili (KP) equations. The DS equations are
the (1+2)-dimensional integrable generalizations of the nonlinear
Schr\"odinger equation  and the KP equations are the same for the
KdV equation. Recently,  interest has appeared	[1-2] into the
(1+2)-dimensional Boiti-Leon-Pempinelli (BLP) equation. Their
integro-differential
equation reduced either to the sine-Gordon or to the sinh-Gordon
equation in a one-dimensional limit. In [1] Boiti, Leon and Pempinelli
used the integrable structure of their equation to extract
a B\"aclund transformation for soliton solution. In [2] it is shown that the
considered equation is hamiltonian
and some soliton's type solutions are constructed.

There are  two purposes of this paper. The first is the study of Darboux and
Laplace transformations to construct  a set of the exact solutions of the
BLP equation. The second is connected with the extremely simple tie between
the BLP equation and the Burgers equation. In the special case
the (1+2) BLP equation is reduced to the (1+1)
Burgers or "anti-Burgers" (i.e., Burgers equation with the time reversed)
equations. Thus we have the interesting question: whether  the BLP equation
has  dissipative solutions in plane? The positive answer is set
forth in the last Section of this paper but here it is necessary to define
more exactly our terminology. We'll do it in the next paragraph.

The Burgers equation is the known model that describes the nonlinear and
dissipative (diffusion) processes. The Burgers equation's solutions we
shall call the {\bf Burgers dissipative structures}. The situation
with the BLP equations is more complex because one is reduced both to the
Burgers and to the anti-Burgers equations. In the Section 5 we show the
way to construct such solutions of the BLP equation which are reduced to the
solutions of the dissipative Burgers equation when  $t\to \infty$.
These solutions we
call the {\bf BLP  dissipative structures}.

The plan of the paper is as follows. In Section 2 we review the BLP equation
(with the reduction to the
Burgers equation) and the BLP's set of discrete symmetries -- the Darboux
(DT) and Laplace (LT) transformations. The result of the multiple DT and
LT is discussed in Section 3. In Section 4 we use the DT for constructing
exact solutions of the BLP equation.  Section 5 is devoted to the Burgers
equation's "dressing procedure" via LT. We leave some final comments to the
last section.
\newline
\newline
{\bf 2. THE BLP EQUATIONS}
\newline
\newline

The integro-differential BLP equation appeared for the first time in [1] and
can be write as the system of two equations ([1], [2]),
$$
\aligned
a_{ty}+(a^2-a_x)_{xy}+2b_{xx}=0,\\
b_t+\left(b_x+2ab\right)_x=0,
\endaligned
\eqno(1)
$$
where $a=a(t,x,y)$, $b=b(t,x,y)$.
We can	introduce new field $p=p(t,x,y)$ which satisfies $b=p_y$ and
rewrite (1) in the following form
$$
\aligned
a_t+2aa_x+(2p-a)_{xx}=0,\\
p_{yt}+\left(p_{yx}+2ap_y\right)_x=0.
\endaligned
\eqno(2)
$$
We'll call (1) and (2) the BLP equations. These equations has $[L,A]$ pair
and the Hamilton structure [2]. On the other hand, the reduction achieved by
imposing $p=0$ maps
the BLP equations (2) onto the differentiated Burgers equation
$$
a_t+2aa_x-a_{xx}=0.
\eqno(3)
$$
If we choose $p=a$ then the eq. (2) reduced "anti-Burgers equation"
which turn into (3) with $t\to-t$, $x\to -x$.

Thus the BLP equations contain the dissipative (Burgers) and anti-dissipative
equation. We
have no  conflict with	this fact and the BLP's  Hamilton structure. It may
be shown that if $b=0$ then the Hamiltonian $H=0$.
Moreover, the fact of Hamiltonian's
existence (and the fact of existence of two additional conservation laws [2])
is right only for
special case asymptotic behavior of $a$ and $b$: $a\to 0$ and $b\to const$ if
$x^2+y^2\to\infty$ [2]. If these conditions are not realized then the
BLP's (anti) dissipative solutions are admissible. We'll demonstrate such
solutions in Sections 4 and 5.

Eq. (1) and (2) is the compatibility condition of the linear system of
equations ($[L,A]$ pair)
$$
\psi_{xy} +a\psi_y+p_y\psi=0,\qquad
\psi_t=\psi_{xx}+2p_x\psi.
\eqno(4)
$$

Let $\phi$ and $\psi$ be solutions of (4). We define  two functions
$\tau=\partial_x(\ln\phi)$ and $\rho=\left(\partial_y(\ln\phi)\right)^{-1}$.
Eq. (4) is covariant with respect to the two types of DT,
$$
\aligned
\psi\to\psi[1]=\rho\psi_y-\psi,\qquad
a\to a[1]=a-\partial_x\ln\rho,\\
\\
p\to p[1]=p-\rho p_y,
\endaligned
\eqno(5)
$$
and
$$
\aligned
\psi\to\psi[1]=\psi_x-\tau\psi,\qquad
a\to a[1]=a-\partial_x\ln(a+\tau),\\
\\
p\to p[1]=p+\tau.
\endaligned
\eqno(6)
$$
In addition, eq. (4) is covariant with respect to the LT
$$
\aligned
\psi\to\psi[1]=\frac{\psi_y}{p_y},\qquad
a\to a[1]=a+\partial_x\ln p_y,\\
p\to p[1]=p+a+\partial_x\ln p_y,
\endaligned
\eqno(7)
$$
and the inverse transformation being given by the formulas
$$
\aligned
\psi\to\psi[-1]=\partial_x\psi+a\psi,\qquad
a\to a[-1]=a-\partial_x\ln\left(p-a\right)_y,\\
p\to p[-1]=p-a.
\endaligned
\eqno(8)
$$

{\bf Remark 1.} The DT (5)-(6) and the LT (7)-(8) are mutually complementary
with each other. It
means that
it is not possible to convert DT into LT and inversely. In fact, for the
DT we get $\psi[1]=0$ if $\phi=\psi$ (see (5)-(6)). But if $\psi[-1]=0$
(see LT (8)) then
we get $p=a$ (see eq. (4)) and $a[-1]$ can't be calculated
from (8). It is easy to verify the analogous conclusion for the LT (7).\newline\newline

{\bf Remark 2.} Let one defines three functions $u$, $v$, $\chi$,
$$
a=-\partial_x\ln u,\qquad b=p_y=-uv,\qquad
\chi=\frac{\partial_x\psi}{u}
$$
and cone variables $\xi$ and $\eta$,
$$
\partial_y=\partial_{\eta}-\partial_{\xi},\qquad
\partial_x=\partial_{\eta}+\partial_{\xi},
$$
then the $L$-equation from (4) can be rewritten in the	Zakharov-Shabat
equation's form
$$
\partial_{\eta}\Phi=\sigma_3\partial_{\xi}\Phi+U\Phi,
\eqno(9)
$$
where
$$
\Phi=\pmatrix
\psi\\
\chi
\endpmatrix,
\qquad
U=\pmatrix
0&u\\
v&0
\endpmatrix,
\qquad
\sigma_3=\pmatrix
1&0\\
0&-1
\endpmatrix.
$$
In Ref. [3], we applied the DT for (9) to construct exact solutions to the
Dirac equations with (1+1) potentials and to the DS equations. It is not
difficult to check that the DT from [3] is the superposition formula for
the two simpler Darboux transformations given by formulas (5) and (6).\newline\newline

{\bf Remark 3.} Eq. (9) is the spectral problem for the DS equations. LT
produce an explicitly invertible B\"acklund autotransformations for the
DS  equations [4]. In Ref. [5] we showed that these transformations
allow one to construct solutions to the DS equations that fall off in all
directions in the plane according to exponential and algebraic law.
\newline
\newline
{\bf 3. CRUM LAW}
\newline
\newline

In Ref. [3], we obtained the extended Crum law for the Eq. (9). In
this Section we will consider this procedure for the DT (5)-(6)
and LT (7)-(8).

Let us consider the DT (5) and define
$Q_{_N}\psi\equiv \partial^{N+1}_y\partial_x\psi$.
It is easy to check that if $\psi$ is a solution of (4) then $Q_{_N}$
can be written as
$$
Q_{_N}=-a\partial^{N+1}_y-(p+Na)_y\partial^N_y+
\sum_{k=1}^{N-1}c_{_{N,k}}\partial^k_y-\partial^{N+1}_y\,p,
\eqno(10)
$$
where $c_{_{N,k}}= c_{_{N,k}}(a,p_y;a_x,p_{xy};
a_y,p_{yy};...)$.

Let  $\left\{\psi,\,\psi_1,\,...\,\psi_{_N}\right\}$ are
$N+1$ particular solutions of (4). After
$N$-time DT (5) we get
$$
\psi[N]=-\psi+\sum_{k=1}^N\alpha_{_{N,k}}\partial^k_y\psi.
\eqno(11)
$$
Plugging (11) into
$$
\psi_{xy}[N] +a[N]\psi_y[N]+p_y[N]\psi[N]=0
\eqno(12)
$$
and using (10) we get
$$
a[N]=a-\partial_x\ln\alpha_{_{N,N}},\qquad
p[N]=p-\sum_{k=1}^N\alpha_{_{N,k}}\partial^{k}_yp.
$$
To compute $\alpha_{_{N,k}}$ we take into account that
$$
\sum_{k=1}^N\alpha_{_{N,k}}\partial^k_y\psi_j=\psi_j,\qquad
j=1,...,N,
$$
therefore
$$
\aligned
\psi[N]=\frac{D_{_N}(1\mid 1)}{D_{_N}(1,N+2\mid 1,2)},\qquad
a[N]=a-\partial_x\ln\frac{D_{_N}(1,2\mid 1,2)}{D_{_N}(1,N+2\mid 1,2)},\\
p[N]=p+(-1)^N\frac{D_{_N}(1\mid 2)}{D_{_N}(1,N+2\mid 1,2)},
\endaligned
\eqno(13)
$$
where $D_{_N}(i,j\mid k,m)$ are determinants which can be obtained
by the deletion of columns $i$, $j$ and rows $k$, $m$ from the
$(N+2)\times (N+2)$ determinant $D_{_N}$
$$
D_{_N}=\vmatrix
\partial^{N+1}p&\partial^{N}p&\partial^{N-1}p&...&\partial p&0\\
\partial^{N+1}\psi&\partial^N\psi&\partial^{N-1}\psi&...&\partial\psi&\psi\\
\partial^{N+1}\psi_1&\partial^N\psi_1&\partial^{N-1}\psi_1&...&\partial\psi_1&\psi_1\\
.&{}&{}&{}&{}&.\\
.&{}&{}&{}&{}&.\\
.&{}&{}&{}&{}&.\\
\partial^{N+1}\psi_{_N}&\partial^N\psi_{_N}&\partial^{N-1}\psi_{_N}&...&\partial\psi_{_N}&\psi_{_N}
\endvmatrix.
\eqno(14)
$$
In (14) for brevity $\partial_y\equiv\partial$.

Now, let us consider the DT (6).
We define
$P_{_N}\psi=\partial^{N+1}_x\partial_y\psi$, then
$$
P_{_N}=-p_y\partial^N_x+\sum_{k=0}^{N-1} \beta_{_{N,k}}\partial^k_x+
\lambda_{_N}\partial_y,
\eqno(15)
$$
As far as
$P_{_{N+1}}=\partial_xP_{_N}$, therefore
$$
\aligned
\beta_{_{N+1,N}}=\beta_{_{N,N-1}}-p_{xy},\qquad
\beta_{_{N+1,k}}=\partial_x\beta_{_{N,k}}+\beta_{_{N,k-1}},\,\,\,1\le k\le N-1,\\
\beta_{_{N+1,0}}=\partial_x\beta_{_{N,0}}-p_y\lambda_{_N},\qquad
\lambda_{_{N+1}}=\partial_x\lambda_{_N}-a\lambda_{_N}.
\endaligned
\eqno(16)
$$
After $N$-time DT (6) we get
$$
\psi[N]=\partial^N_x\psi+\sum_{k=0}^{N-1}\alpha_{_{N,N-k}}\partial^k_x\psi.
\eqno(17)
$$
Substituting (17) into (12) and taking (15), (16) into account we get
$$
\aligned
a[N]p_y[N]=ap_y+
\partial_y\left(\partial_x(\alpha_{_{N,1}}-Np)+\alpha_{_{N,2}}-\frac{1}{2}\alpha_{_{N,1}}^2\right),\\
p[N]=p-\alpha_{_{N,1}}.
\endaligned
\eqno(18)
$$
To compute $\alpha_{_{N,k}}$ we take into account that ($j=1,...,N$)
$$
\partial^N_x\psi_j+\sum_{k=0}^{N-1}\alpha_{_{N,N-k}}\partial^k_x\psi_j=0,
\eqno(19)
$$
therefore
$$
\psi[N]=\frac{W_{_N}}{W_{_N}(1\mid 1)},\qquad
\alpha_{_{N,1}}=-\frac{W_{_N}(2\mid 1)}{W_{_N}(1\mid 1)}, \qquad
\alpha_{_{N,2}}=\frac{W_{_N}(3\mid 1)}{W_{_N}(1\mid 1)},
\eqno(20)
$$
where
$$
W_{_N}=\vmatrix
\partial^N\psi&\partial^{N-1}\psi&...&\psi\\
\partial^N\psi_1&\partial^{N-1}\psi_1&...&\psi_1\\
.&{}&{}&.\\
.&{}&{}&.\\
.&{}&{}&.\\
\partial^N\psi_{_N}&\partial^{N-1}\psi_{_N}&...&\psi_{_N}
\endvmatrix
\eqno(21)
$$
and $\partial_x\equiv\partial$.

With the help of (7) and (8) one can find now the result of $N$-fold LT
$a[N]$ ($a[-N]$) and $b[N]$ ($b[-N]$) (it is convenient to use fields $b$
instead of fields $p$ in these cases).
Let us consider the LT (7). After $N$-time transformations we get for
the $\psi[N]$ (compare with the (11))
$$
\psi[N]=\sum_{k=1}^N\alpha_{_{N,k}}\partial^k_y\psi.
\eqno(22)
$$
It is convenient to redesignate  $a[N]\equiv a_N$, $b[N]\equiv b_N$.
Substituting (22) into (12) we get
$$
a_N=a-\partial_x\ln \alpha_{_{N,N}},\qquad
b_N=b+\partial_y\left(Na-\partial_x\ln (\alpha_{_{N-1,N-1}}
\alpha_{_{N,N}})\right)
\eqno(23)
$$
with the supplementaty condition
$$
\sum_{k=1}^N \alpha_{_{N,k}}\partial^{k-1}_yb=0.
\eqno(24)
$$

Thus it is necessary to find  coefficients  $\alpha_{_{N,k}}$
and to use (22)-(23). After the calculation we get
$$
\alpha_{_{N,N}}=\left(\prod_{j=1}^N\,b_{j-1}\right)^{-1},
\eqno(25)
$$
so (23) is the nonlinear superposition formula for the BLP equations.
To compute $\psi[N]$ we must  obtain the rest coefficients
$\alpha_{_{N,k}}$. To do it we introduce the operators
$$
{\bold f}_j=\frac{1}{b_j}\partial_y,\qquad
{\bold F}={\bold f}_{_{N-1}}{\bold f}_{_{N-2}}...{\bold f},
$$
where $j=0,...,N-1$, $b_0=b$, ${\bold f}_0={\bold f}$. It's clear that
$$
\psi[N]= \sum_{k=1}^N \alpha_{_{N,k}}\partial^k_y\psi={\bold F}\psi.
\eqno(26)
$$
The coefficients  $\alpha_{_{N,_k}}$ from the (26) can be calculated
by the Crum law's manner if to find $N$ functions $\theta_j$ ($j=1,...,N$)
such that ${\bold F}\theta_j=0$. It is easy to obtain these functions
in the following way: for an arbitrary index $j$ we construct the sequence
of equations for the $j$ functions $\theta_j^{(k)}$ ($k=1,...,j$),
$$
{\bold f}_{k-1}\theta_j^{(k-1)}=\theta_j^{(k)},\qquad
\theta_j^{(0)}=\theta_j,\qquad \theta_j^{(j-1)}=1,\qquad
\eqno(27)
$$
The system (27) can be solved and we get $N$ functions $\theta_j$,
$$
\theta_j=<b<b_1<...<b_{j-1}>...>_y.
$$
In what follows $<...>_{x,y}$ denote the integration by the $x,\,y$-variables,
$$
<S>_x=\frac{1}{2}\int dz\,sgn(x-z) S(z,y,t),\qquad
<S>_y=\frac{1}{2}\int dz\,sgn(y-z) S(x,z,t).
$$

We express the $\alpha_{_{N,1}}$ from (24), then
$$
{\bold F}\theta_j=\sum_{k=2}^N M_{kj}\alpha_{_{N,k}}=0,\qquad
M_{kj}=\partial^k_y\theta_j-\frac{\partial^{k-1}_yb}{b}\partial_y\theta_j.
\eqno(28)
$$
It is obvious that $M_{k1}=M_{k2}=0$. Plugging (25) into (28) we get the
nonhomogeneous system of linear equations (for desired coefficients)
which can be solved by the
Kramer's formulae.

The result of $N$-fold LT (8) can be obtained by analogy with LT (7).
It is easy to see that $\psi[-N]$ has the form (17) with new coefficients
$\alpha_{_{N,k}}$. It is means that
$b[-N]$ and $a[-N]$ can be computed by the (18). Therefore our task is reduced
to the search of new coefficients $\alpha_{_{N,k}}$. These coefficients
may be expressed in terms of determinants (20), (21) with the change
$\psi_k\to \theta_k$ where these functions ($\theta_k$)
satisfy the equation (19).

To find $\theta_k$ we rewrite (19) as
$$
\left(\partial_x+a_{_{N-1}}\right)\left(\partial_x+a_{_{N-2}}\right)...
\left(\partial_x+a\right)\theta_j=0,\qquad j=1,...,N
\eqno(29)
$$
where we designate $a[-k]\equiv a_k$, $b[-k]=b_k$ for the convenience.
The (29) can be rewritten as a system,
$$
\cases \left(\partial_x+a_j\right)\theta_m^{(j)}=\theta_m^{(j+1)},&\text
{for $j=0,...,m-2,$}\\
\left(\partial_x+a_{m-1}\right)\theta_m^{(m-1)}=0,\endcases
\eqno(30)
$$
for  any $\theta_m$. In (30) $\theta_m^{(0)}=\theta_m$, $m=2,...,N$, and
$$
\theta_1=\exp\left(-<a>_x\right).
$$
By solving (30) we get the required functions
$$
\theta_m=\theta_1\,\Phi_m,\qquad
\Phi_m=<b_1<b_2<...<b_{m-1}>...>_x.
$$
\newline
\newline
{\bf 4. SOLITONS ON THE VACUUM BACKGROUND}
\newline
\newline

In Ref. [2], author applied the B\"acklund transformations to construct
singular solutions that fall off according to rational law in all
directions in the plan with $a\to 0$ and $b\to const$
if $x^2+y^2\to\infty$. In this Section we show that the
DT (5), (13) and (6), (18) allow one
to obtain a rich set of exact solutions of the BLP equations (2).
We'll consider two examples:\newline
(i) the localized nonsingular solution $b$
falling off according to the exponential law as a function of $x$
and according to the rational law as a function of $y$;\newline
(ii) the "blow-up"  solution which is nonsingular when $t>0$ and which has
singularity when $t<0$ (may be it is more correctly to call such solutions
"anti-blow-up" or "white hole" solutions).

Let $a=b=0$. To construct the solution (i) we choose the solution of (4)
in the form
$$
\phi=\exp\left(\mu^2t\right)\cosh(\mu x)+B(y),
$$
where $B=B(y)$ is an arbitrary differentiable function.
Taking (6) into account we get
$$
\aligned
b[1]=-\frac{\mu B'\sinh(\mu x)\exp\left(\mu^2 t\right)}
{\left(B+\cosh(\mu x)\exp(\mu^2 t)\right)^2},\qquad
a[1]=-\frac{\mu(\exp\left(\mu^2 t\right)+B\cosh(\mu x))}
{\sinh(\mu x)\left(B+\cosh(\mu x)\exp(\mu^2 t)\right)}.
\endaligned
\eqno(31)
$$
If we choose the  function $B(y)$ increasing according to the
exponential law ($y\to\pm\infty$) that the solution (31) has nonclosed level
curves, so we choose one as an even polynomial (the property of being even
guarantees the nonsinular behavior of (31)). Thus, assuming
$B(y)=\nu^2 y^{2N}+\lambda^2$ ($\nu$ and $\lambda$ are const and
$\Im\nu=\Im\lambda=0$)
we get the desired solution
$$
b[1]=-\frac{2\mu N\nu^2 y^{2N-1}\sinh(\mu x)\exp\left(\mu^2 t\right)}
{\left(\lambda^2+\nu^2 y^{2N}+\cosh(\mu x)\exp(\mu^2 t)\right)^2}.
\eqno(32)
$$
The regular solution (32) falling off according to the exponential law
when $x\to\pm\infty$
and according to the rational law  ($1/y^{2N+1}$)
when  $y\to\pm\infty$, at the same time the singular solution $a[1]$ has level
curves along $y$-line.

To construct (anti) blow-up solutions of BLP (ii) we shall apply two DT (5) and (6)
on the vacuum background ($a=b=0$). As a result we get
$$
b[2]=\frac{W_{_2}(1\mid 1)\,D_{_2}(1,4\mid 1,2)}
{D^2_{_2}(1,1\mid 1,1)},
\eqno(33)
$$
Two functions $\psi_k$ ($k=1,2$) from the determinants
$D$ and $W$  (see (14) and (21)) are the solutions of (4). We
choose ones in the form
$$
\psi_1=C_1\exp\left(\lambda^2 t\right)\cosh(\lambda x)+
\cosh(\alpha y),\qquad
\psi_2=C_2\exp\left(-\mu^2 t\right)\sin(\mu x)+\sinh(\beta y).
$$
Then $D_{_2}(1,1\mid 1,1)=D_1(y)+D_2(x,y,t)$
where
$$
\aligned
D_1(y)=\beta\cosh(\beta y)\cosh(\alpha y)-
\alpha\sinh(\beta y)\sinh(\alpha y),\\
D_2(x,y,t)=\beta C_1\exp\left(\lambda^2 t\right)\cosh(\lambda x)\cosh(\beta y)-
\alpha C_2\exp\left(-\mu^2 t\right)\sinh(\alpha y)\sin(\mu x).
\endaligned
$$
If $\beta>\alpha>0$ then $D_1(y)>0$ for the $y \in(-\infty,+\infty)$.

Let us consider $D_2(x,y,t)$. It is clear that if $C_1>0$ then
$D_2(x,0,t)>0$. So the solution (33) is nonsingular by the $y=0$.
A singularity can appear if $D_2(x,y,t)<0$. It is possible if such
values $x'$, $y'$, $t'$ of variables $x$, $y$, $z$ exist that
$D_2(x',y',t')=0$, i.e.
$$
\rho\exp\left((\lambda^2+\mu^2)t')\right)\cosh(\beta y')\cosh(\lambda x')=
\sinh(\alpha y')\sin(\mu x'),
\eqno(34)
$$
when $\rho\equiv \beta C_1 (\alpha C_2)^{-1}$. It is obvious that the
condition (34) is not right by the $y'=0$. We shall change (34) on,
$$
\rho\exp\left((\lambda^2+\mu^2)t')\right)\mid\theta_1(y')\mid=
\mid\theta_2(x')\mid,
\eqno(35)
$$
with
$$
\theta_1(y)=\frac{\cosh(\beta y)}{\sinh(\alpha y)},\qquad
\theta_2(x)=\frac{\sin(\mu x)}{\cosh(\lambda x)}.
$$
We are interested in selection of parameters such that (35) is not right
because this choice guarantees the nonsingular behavior of (33). Let
$y_0$ and $x_0$ are solutions of  equations
$$
\tanh(\beta y_0)\tanh(\alpha y_0)=\frac{\alpha}{\beta},\qquad
\tanh(\lambda x_0)\tan(\mu x_0)=\frac{\mu}{\lambda},
$$
and $x_0$ is the maximum point of $\theta_2(x)$. Since
$\exp\left((\lambda^2+\mu^2)t')\right)\ge 1$ by the $t'\ge 0$ then for
the such values of $t$ we can to choose
$$
\rho>\frac{\theta_2(x_0)}{\theta_1(y_0)}.
\eqno(36)
$$
Thus the condition (36)  guarantees the nonsingular behavior of (33)
by the $t\ge 0$.  It is possible to choose parameters such that
(33) is singular (blow-up) solution in a defined region and by
the $t<0$.

The solution (33) has oscillation by the variable $x$. We can
obtain a blow-up solution without one. To construct it we choose
$\psi_2$ as
$$
\psi_2=C_2\exp\left(\mu^2 t\right)\sinh(\mu x)+\sinh(\beta y).
$$
The solution (33) will be regular by the $t\ge 0$ (and has blow-up in a
defined bounded region by the $t<0$ and by the special choice of parameters)
if $\lambda>\mu$ and
$$
\rho>\frac{\tilde{\theta}_2(\tilde{x_0})}{\theta_1(y_0)},\qquad
\tilde{\theta}_2(x)=\frac{\sinh(\mu x)}{\cosh(\lambda x)},
$$
where $\tilde{x_0}>0$ is maximum point of the function	$\tilde{\theta}_2(x)$.
\newline
\newline
\newline
{\bf 5. THE BURGERS EQUATION'S DRESSING PROCEDURE}
\newline
\newline
In Section 2 we showed that the reduction achieved by imposing $p=0$ maps
the BLP equations (2) onto the differentiated Burgers equation (3).
This fact allows us to use solutions of the Burgers equations in the capacity
of the "supplier" of the BLP's solutions.
Namely, using an arbitrary solution of the Burgers equation which has
parameters depending on $y$ it is possible to construct exact solutions
of (2) with the help of the LT (8) (we can't apply the LT (7) by the
$p=0$).

It is convenient to redefine variables by the following way
$$
x=\frac{\xi}{\sqrt{\nu}},\qquad y=\frac{\eta}{\nu\sqrt{\nu}},\qquad
a=\frac{A(\xi,\eta,t)}{2\sqrt{\nu}},\qquad
p=\frac{P(\xi,\eta,t)}{\nu\sqrt{\nu}}.
$$
Here $\nu>0$ and $\mu$ is arbitrary constant. As a result the BLP equations
take the form
$$
\aligned
A_t+AA_{\xi}-\nu A_{\xi\xi}=
-4 P_{\xi\xi},\\
P_{t\eta}+
\left(AP_{\eta}+\nu P_{\xi\eta}\right)_{\xi}=0.
\endaligned
\eqno(37)
$$
$\nu$ is the parameter that may be called the coefficient of viscosity. If
we shall assume that $P=0$ and $A$ is a solution of the Burgers
equation that (8) allows us to construct  solutions of the BLP
equations (37),
$$
P[-1]=-\frac{\nu}{2}A,\qquad
A[-1]=A-2\nu\partial_{\xi}\ln\left(A_{\eta}\right).
\eqno(38)
$$
This is the procedure that we call the the Burgers
equation's "dressing procedure". Let us consider a simple example
for the shock wave of the Burgers equation [6],
$$
A=A_1+\frac{A_2-A_1}{1+\exp\left\{\frac{A_2-A_1}{2\nu}(\xi-Ut)\right\}},
$$
where
$$
A_1=A_1(\eta),\qquad A_2=A_2(\eta),\qquad U=\frac{A_1+A_2}{2}.
$$
Assuming that $A_1=0$ and taking (38) into account we get
$$
\aligned
A[-1]=\frac{A_2\left[G^2A_2(\xi-A_2 t)-2\nu(2G^2+3G+1)\right]}
{(1+G)\left[G A_2(\xi-A_2 t)-2\nu(1+G)\right]},\\
P[-1]=-\frac{\nu A_2}{2(1+G)},\\
G\equiv\exp\left\{\frac{A_2(2\xi-A_2 t)}{4\nu}\right\}.
\endaligned
\eqno(39)
$$
It is interesting to consider the behavior of (39) by the $\nu\to 0$.
We suppose that $A_2>0$ for all values of the $\eta$. Then for the
$2\xi\ne A_2 t$, we get
$$
P[-1]\to 0,\qquad A[-1]\to A_2,
$$
and for the $2\xi=A_2 t$
$$
P[-1]\to 0 ,\qquad
A[-1]\to \frac{A_2}{2}.
$$
Using other well known solutions of the Burgers equation (see for example the
remarkable monograph [6]) it
is easy to construct a rich set of the exact solutions of the
BLP equations via  the Burgers equation's "dressing procedure". Let $A$ is
the the dissipative solution of the Burgers equations, i.e. $A\to 0$ if $t\to
\infty$. Using the (38) we get dissipative solution $P[-1]$ therefore
we get a tending to zero (when $t\to \infty$) right term in the first
equation (37) and  $A[-1]\to A_B$, where $A_B$ is solution of the
Burgers equation. We call such BLP's solutions the  BLP
dissipative structures (see Section 1).

For example we consider the dissipative solution of the BLP equations,
$$
A=-2\nu\partial_{\xi}\ln f,\qquad
f=\alpha+\sqrt{\frac{\beta}{t}}\exp\left(-\frac{\xi^2}{4\nu t}\right),
\qquad
P=0,
\eqno(40)
$$
where $\alpha=\alpha(y)$ and $\beta=\beta(y)$ are some arbitrary
functions (if $\alpha$ and $\beta$ are constants then (40) is the
so called N-waves
solution of the dissipative Burgers equation [6]).
Substituting (40) into (38) we get
$$
A[-1]=-\frac{2\nu}{\xi}+\frac{\alpha\xi}
{tf},\qquad
P[-1]=-\frac{\nu\xi(f-\alpha)}{2tf}.
$$
It easy to see that if $t\to+\infty$ then
$$
P[-1]\to 0,\qquad
A[-1]\to A_B=-\frac{2\nu}{\xi}+\frac{\xi}{t},
$$
and $A_B$ is the solution of the  Burgers equation.
\newline
\newline
\newline
\newline
{\bf 6. CONCLUSION}
\newline
\newline

In this paper we have studied in detail the Darboux and Laplace
transformations for the BLP equations. These transformations represents
a really fruitful mathematical tool in the search for exact solutions
of  integrable nonlinear evolution equations
in two spatial dimensions. The physical interpretation
of the BLP equations within the framework of  theory of dissipative
processes is not entirely clear to me because the BLP equation reduced to a
differented Burgers equation and to a anti-Burgers equation too.
Therefore only	parth of BLP exact solutions has  dissipative behavior.
In this work we have showed how these solutions
(BLP  dissipative structures) associated with the dissipative Burgers
equation may be obtained.
\newline
\newline
{\bf Acknowledgement}
\newline
\newline

It is pleasure to thank Dr. S.B. Leble and anonymous referee for the valuable
comments.
This work was partially sponsored by the RFFI under Grant No. 96-01-01789,
96-01-01408.

\enddocument